\documentclass[preprint]{aastex}
\usepackage{graphicx}
\usepackage{subfigure}
\usepackage{mathrsfs,amssymb}
\usepackage{amsmath}
\usepackage{natbib}
\usepackage{color}
\usepackage{ulem}
\usepackage{bm}

\def\gtsima{$\; \buildrel > \over \sim \;$}
\def\ltsima{$\; \buildrel < \over \sim \;$}
\def\gsim{\lower.5ex\hbox{\gtsima}}
\def\lsim{\lower.5ex\hbox{\ltsima}}
\def\simgt{\lower.5ex\hbox{\gtsima}}
\def\simlt{\lower.5ex\hbox{\ltsima}}
\def\simpr{\lower.5ex\hbox{\prosima}}
						
\begin{document}

\title{{\Large Infrared background signatures of the first black holes}} 

\author{
Bin Yue\altaffilmark{1},
Andrea Ferrara\altaffilmark{2},
Ruben Salvaterra\altaffilmark{3},
Yidong Xu\altaffilmark{1},
Xuelei Chen\altaffilmark{1,4}
}
\altaffiltext{1}{National Astronomical Observatories, Chinese Academy of Sciences,
20A Datun Road, Chaoyang, Beijing 100012, China}
\altaffiltext{2}{Scuola Normale Superiore, Piazza dei Cavalieri 7, I-56126 Pisa, Italy}
\altaffiltext{3}{INAF, IASF Milano, via E. Bassini 15, I-20133 Milano, Italy}
\altaffiltext{4}{Center of High Energy Physics, Peking University, Beijing 100871, China}

\begin{abstract}
Angular fluctuations of the Near InfraRed Background (NIRB) intensity are observed up to scales 
$\simlt 1^{\ensuremath{^{\circ}}}$. 
Their interpretation is challenging  as even after removing the contribution from detected sources,
the residual signal is $>10$ times higher than expected from distant galaxies below the detection limit
and first stars. We propose here a novel interpretation in which early, intermediate mass, accreting direct
collapse black holes (DCBH),  which are too faint to be detected individually in current surveys,
could explain the 
observed fluctuations. We find that a population of highly obscured ($N_{\rm H}\simgt 10^{25}~\rm cm^{-2}$) 
DCBHs formed in metal-free halos with virial temperature $10^4$~K at $z\simgt 12$, can explain the observed
level $\approx 10^{-3}$ (nW m$^{-2}$ sr$^{-1})^2$ of the 3.6 and 4.5 $\mu$m fluctuations on scales $>100''$. 
The signal on smaller scales is instead produced by undetected
galaxies at low and intermediate redshifts. Albeit Compton-thick, at scales
$\theta> 100''$ DCBHs produce a CXB (0.5-2 keV)-NIRB ($4.5 \rm \mu m$) cross-correlation signal of $\simeq 10^{-11}$ erg s$^{-1}$ cm$^{-2}$ nW m$^{-2}$ sr$^{-1}$ slightly dependent on the specific value of the absorbing gas
column  ($N_{\rm H} \approx 10^{25}~\rm cm^{-2}$) adopted and in agreement with the recent
measurements by \cite{2012arXiv1210.5302C}.
At smaller scales the cross-correlation is dominated by the emission 
of high-mass X-ray binaries (HMXB) hosted by the same low-$z$, undetected galaxies accounting for small scale 
NIRB fluctuations. These results outline the great potential of the NIRB as a tool to investigate the nature of the first 
galaxies and black holes.
\end{abstract}


\maketitle

\section {INTRODUCTION}\label{intro}
  
Measurements of the Near InfraRed Background (NIRB) 
have raised the hope to open a new window on high redshift galaxies or even 
the first stars 
\citep{2002ApJ...579L..53K,2002MNRAS.336.1082S,2003MNRAS.339..973S,2004ApJ...606..611C,
2004ApJ...608....1K}. 
The observed NIRB fluctuations are largely dominated by low redshift galaxies which are either 
resolved individually or have their signal constrained by the luminosity functions obtained from deep
multiband surveys \citep{2012ApJ...752..113H}. However, even after subtraction of these contributions 
from the images, the majority of the residual signal is still unlikely to come from high redshift galaxies
or first stars \citep{2006MNRAS.367L..11S,2012ApJ...756...92C, 2013MNRAS.431..383Y}. As the fluctuation
measurements have been confirmed by different experiments 
\citep{2005Natur.438...45K,2007ApJ...654L...5K,	
2007ApJ...657..669T,2011ApJ...742..124M,2012ApJ...753...63K,2012Natur.490..514C}, 
the problem remains of what sources produce the measured signal (see \citealt{2012Natur.490..514C}
for an alternative explanation). 

According to the most popular cosmological theory, the first stars formed at $z = 20-30$ 
\citep{2003Sci...300.1904M} in mini-halos (dark matter structures with virial temperatures 
$T_{\rm vir}  = 10^{3-4}$~K) cooling their primordial (metal-free) gas via $\rm H_2$ line emission. 
For a sufficiently intense Lyman-Werner band (LW  $\equiv 11.2-13.6$ eV) UV radiation field, 
$\rm H_2$ can be photo-dissociated (see e.g., \citealt{2010MNRAS.402.1249S}) 
and cooling and star formation are quenched.
Larger halos ($T_{\rm vir} \gtrsim 10^4$~K) do not rely on $\rm H_2$, as hydrogen Ly$\alpha$ line 
emission  and
other processes sustain an almost isothermal collapse preventing gas fragmentation into smaller sub-units.
Under these conditions, theoretical works \citep{2002ApJ...572L..39S,2003ApJ...596...34B,2004MNRAS.354..292K,
2006MNRAS.370..289B,2008MNRAS.383.1079V,2009MNRAS.396..343R,2010MNRAS.402.1249S}
show that the most likely outcome is a rapid ($\approx 1 $ Myr) formation of a direct collapse black hole (DCBH) 
of  mass $10^{4-6} M_\odot$ 
\citep{2003ApJ...596...34B,2006MNRAS.371.1813L,2006MNRAS.370..289B,2009MNRAS.396..343R,
2012ApJ...750...66J,2012arXiv1211.0548J}.  
The halo gas presumably continues to fall onto the
BH thus powering its luminosity at the Eddington rate for 10-100
Myr. These DCBHs are very likely Compton-thick (implying H column densities $N_{\rm H} \gtrsim 10^{25}~\rm cm^{-2}$),
as the gas infall rate is very high and the stacked gas very compact. Compton-thick quasars are heavily absorbed from UV (above 13.6 eV) to X-ray energies, as ionizing photons are absorbed 
by neutral gas surrounding the BH
\citep{2007A&A...463...79G,2009MNRAS.397.1549M}  and reprocessed into
optical-UV ``nebular'' emission \citep{2006ApJ...646..703F}.  
Therefore, they do not contribute significantly to the reionization of the
intergalactic medium.  This rises the hope to explain the observed amplitude of the 
source-subtracted NIRB fluctuations by DCBHs, without overshooting the experimental
constraints on the unresolved fraction of the Cosmic X-ray Background (CXB)
\citep{2012A&A...545L...6S} and reionization. 

Intermediate mass BH seeds such as those formed via the above mechanism, ease 
the problem of explaining the inferred masses, $\sim10^9~M_\odot$, of supermassive BHs 
(SMBH; \citealt{2012MNRAS.422.1690P}) already in place at
$z= 6-7$ \citep{2001AJ....122.2833F}. In fact, starting from a  smaller, stellar-size mass BH, 
accretion should proceed at the Eddington limit throughout the entire Hubble time to build up 
such a large mass.
This seems unlikely. 

The layout of the paper is as follows. In Section \ref{method} we present the spectrum of a Compton-thick BH, 
and introduce the calculation of their contribution to the NIRB and the CXB.
In Section \ref{results} we present our results: the DCBH parameters that fit the measured 
NIRB fluctuations; the predictions of CXB angular power spectrum and the CXB-NIRB cross-correlation.
Conclusions are presented in Section \ref{conclusions}.
We discuss the possibility of the formation of a large number of DCBHs in Appendix \ref{discussions}.

Throughout this paper, we use the same cosmological parameters as in
\cite{2011MNRAS.414..847S}:  $\Omega_m$=0.26, $\Omega_\Lambda$=0.74,
$h$=0.73, $\Omega_b$=0.041, $n=1$ and $\sigma_8$=0.8.
The transfer function is from \cite{1998ApJ...496..605E}. Magnitudes
are given in the AB system.

\section{Model description}\label{method}

\subsection{Spectrum of a Compton-thick accreting BH}\label{spectrum}

The primary emission spectrum of an accreting BH can be described as the sum
of 
three components:
\begin{equation}
L_\nu = L^{\rm MBB}_\nu+L^{\rm PL}_\nu+L^{\rm refl}_\nu,
\label{Lnu}
\end{equation}
namely a multicolor black body spectrum, i.e.,
a combination of black body spectra with different temperatures coming
from different parts of the accretion disc; a power law spectrum 
from a surrounding hot corona; and a reflection component respectively.
The first two components have comparable luminosity \citep{2004IJMPD..13....1M,2005MNRAS.363.1069K}.
The temperature of the disc decreases from inside out and is maximum in the
innermost regions 
with \citep{2000ApJ...535..632M,2005MNRAS.362L..50S}
\begin{equation}
T_{\rm max}=\left(\frac{M_{\rm BH}}{M_\odot}\right)^{-0.25}~\rm keV,
\end{equation}
where $M_{\rm BH}$ is 
the BH mass. This approximation is valid in the case the central object is a Schwarzschild BH, the innermost radius is $\approx$5 times the Schwarzschild radius, and the accretion reaches the Eddington limit.
The spectral energy distribution is then \citep{1984PASJ...36..741M}
\begin{equation}
L^{\rm MBB}_\nu=L_{\rm MBB}\int_{0}^{T_{\rm max}}B_{\nu}(T)\left(\frac{T}{T_{\rm max}}\right)^
{-11/3}\frac{dT}{T_{\rm max}},
\end{equation}	
where $B_{\nu}(T)$ is the emission spectrum of a black body with temperature $T$,
$L_{\rm MBB}$ is a factor used for normalization.

The hot corona emission spectrum is usually parametrized as a power law with an exponential cut-off, i.e. 
\begin{equation} 
L^{\rm PL}_{\nu} = L_{\rm PL} \nu^{-\alpha_s}{\rm exp}(-h_p\nu/E_{\rm cut}), 
\end{equation}
where $h_p$ is the Planck constant. We adopt $\alpha_s = 1$ and $E_{\rm cut} = 300 $~keV 
\citep{2004MNRAS.347..144S}. Again $L_{\rm PL}$ is the normalization factor.
As in \citet{2005MNRAS.362L..50S}, the power-law is truncated below the peak 
of the disc component, i.e. $\sim 3T_{\rm max}$. 

A fraction of the radiation emitted by the hot corona is reflected by the disc,
and must be added to the original spectrum.
During this process, high energy photons would be Compton scattered to lower energies,
producing a ``Compton hump" at energy around 30~keV \citep{2007ApJ...657..448F}.
We use the {\tt pexrav} \citep{1995MNRAS.273..837M} package 
included in {\tt xspec}\footnote{http://heasarc.nasa.gov/xanadu/xspec/} to 
calculate this reflection component, 
$L^{\rm refl}_\nu$,
adopting an angle between the normal to the disc and 
the line-of-sight of $60^{\ensuremath{^{\circ}}}$. The
reflection scaling factor, i.e. the solid angle (in units of $2\pi$) subtended by the disc as viewed 
from the X-ray source, is set equal to 1, which corresponds to the case of an isotropic source 
located above the disk (see {\tt xspec} manual for more details).
As DCBHs form in pristine gas, 
we adopt a zero gas metallicity. 

\begin{figure}
\begin{center}
\centering{
\subfigure{\includegraphics[scale=0.4]{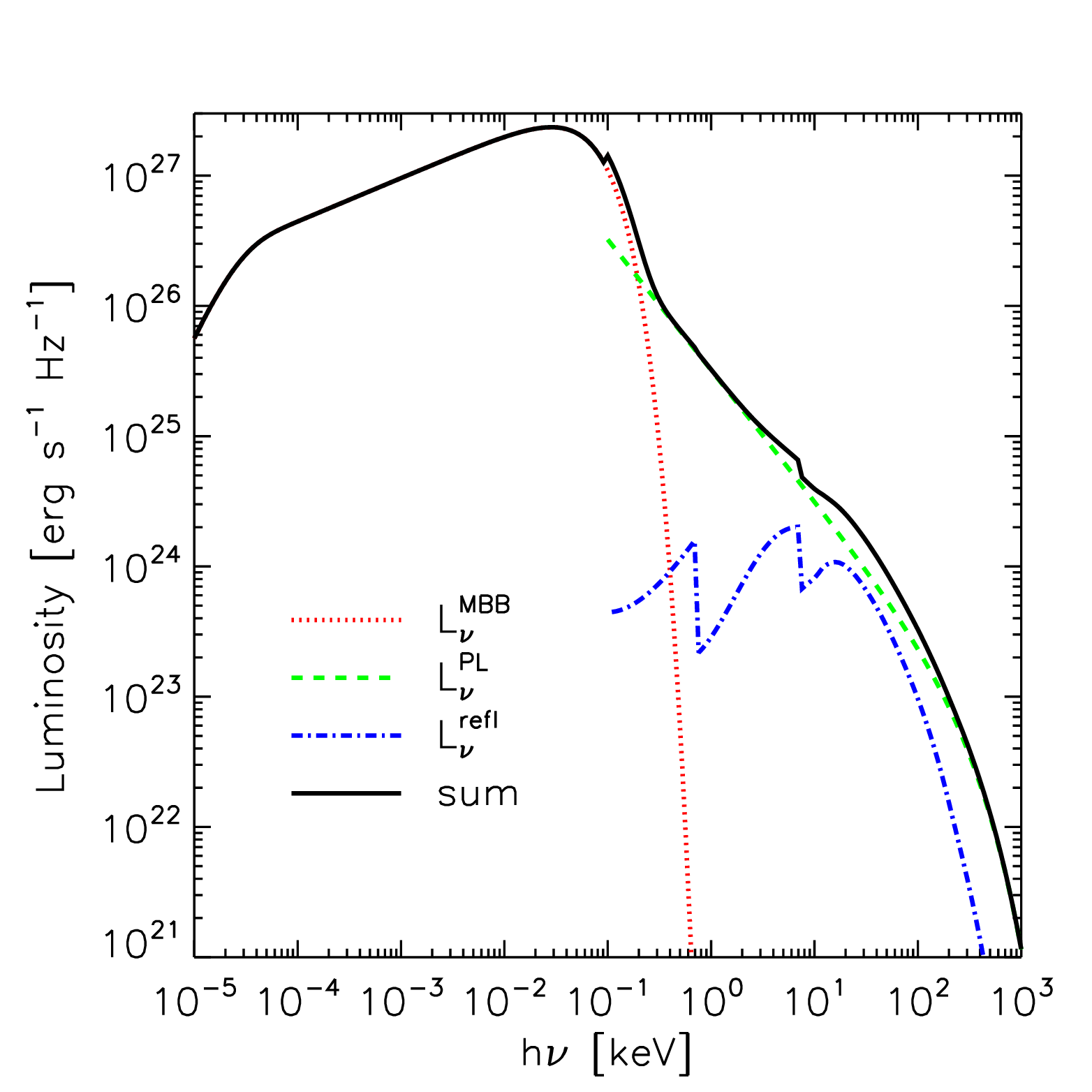}}
\subfigure{\includegraphics[scale=0.4]{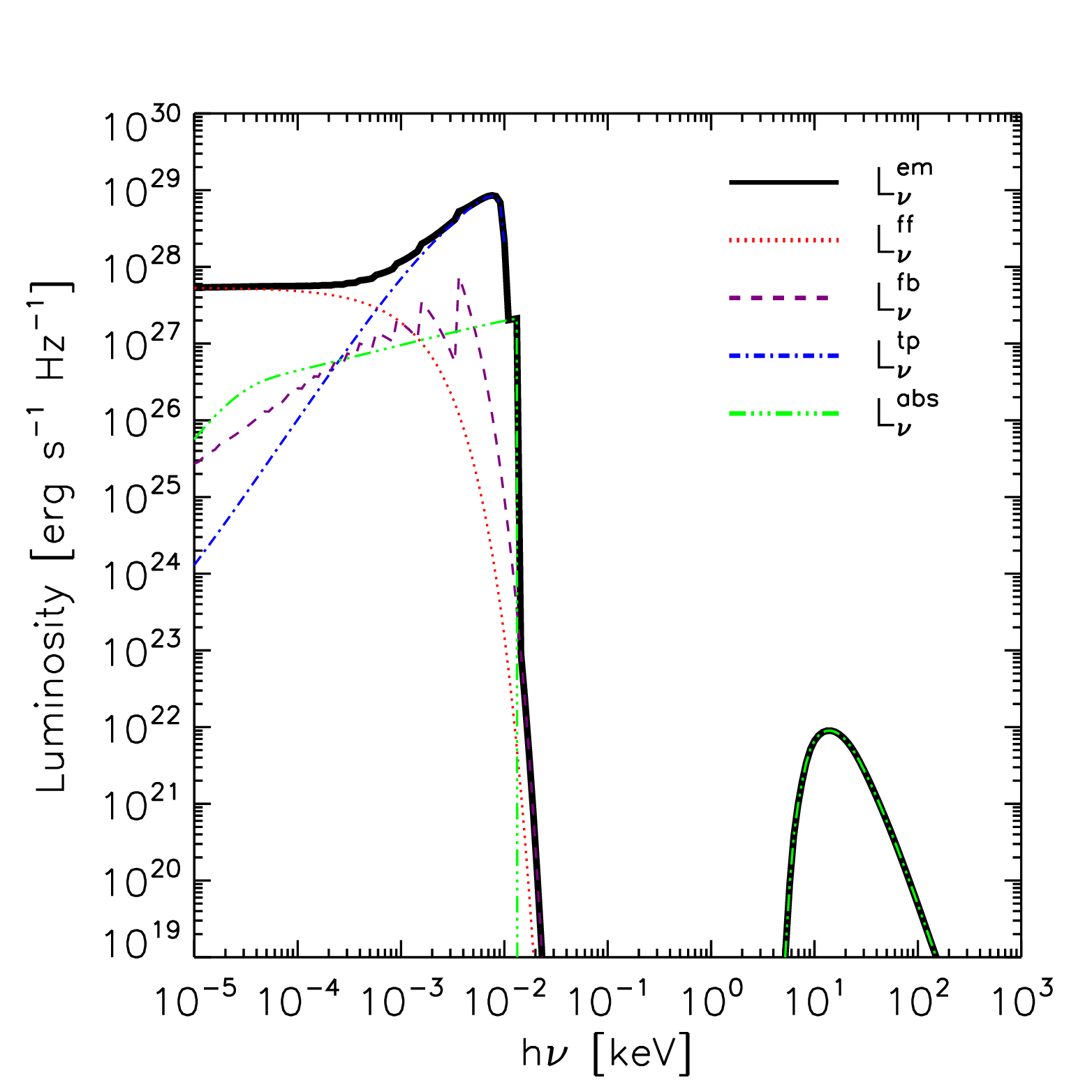}}
\caption{ 
{\textit Left}: the primary spectrum (solid) 
for a BH with $M_{\rm BH}=10^6~M_\odot$
and its three components.
{\textit Right}: the emerging (thick solid line) quasar spectrum
of above BH when $N_{\rm H} = 1.5\times10^{25}~\rm cm^{-2}$
and the four components (thin lines).
\label{LQSO}
}
}
\end{center}
\end{figure}

Throughout this paper we assume that the BH radiates at the Eddington limit
when accreting, so that
its bolometric luminosity is $L_{\rm Edd} = 1.3\times10^{38}M_{\rm BH}~\rm
erg\,s^{-1}$,
as done in many previous works 
(e.g., \citealt{2004ApJ...604..484M,2004MNRAS.351L..71C,2009ApJ...701L.133A,
2009ApJ...694..842M, 
2011ApJ...739....2P}). However, our results do not depend critically
on this assumption as a smaller Eddington ratio can be compensated 
(up to a certain point) by 
longer accretion times.
$L_{\rm MBB}$ and $L_{\rm PL}$ are then determined by 
$\int L^{\rm MBB}_\nu d\nu = \int L^{\rm PL}_\nu d\nu$ and 
$\int (L^{\rm MBB}_\nu +L^{\rm PL}_\nu +L^{\rm refl}_\nu)d\nu = L_{\rm Edd}$.
As an example, we plot the final primary spectrum of a BH with mass $10^6~M_\odot$
and its three components in the upper panel of Fig. \ref{LQSO}.
The reflection spectrum only contributes $\approx10\%$ of the total energy. 

As DCBHs are enshrouded by a massive accreting envelope, it is very
likely that they are Compton-thick. In this case, the emerging
spectrum, filtered by such medium, will be quite different from the primary one. 
In fact, most of the the photons above 13.6~eV are either  absorbed or
scattered. We assume that the absorbing gas distribution is spherically
symmetric, so that also the reflection component is absorbed.
This is motivated by the fact metal-free gas disks are relatively hot and in $T_{\rm vir} > 10^4$~K halos 
are likely to be fat \citep{2005ApJ...633..624V}.
With the primary spectrum expressed by Eq. (\ref{Lnu}),
we follow the \citet{1997ApJ...479..184Y} (Y97) to calculate the absorbed and scattered spectrum, 
$L^{\rm abs}_\nu$,
despite of possible uncertainties arising when $N_{\rm H} > 5\times10^{24}~\rm cm^{-2}$.
A brief introduction of the method and formulae is given in Appendix \ref{Y97},  
we refer interested readers to the original paper for more details.
The $L^{\rm abs}_\nu$ of a $10^6~M_\odot$ BH 
with $N_{\rm H} = 1.5\times10^{25}~\rm cm^{-2}$ is shown in the bottom 
panel of Fig. \ref{LQSO}.

The energy of photons is transferred to 
electrons
by both photoelectric absorption and 
Compton scattering.
While each UV photon
typically ionizes only one neutral atom, we have to take into account
that each X-ray photon generates a high energy electron 
directly,	
which in turn initiates a collisional ionization
cascade. The total number of ionizations per unit time is
\begin{equation}
Q_{\rm H}=Q_{\rm H,UV}+Q_{\rm H,X},
\label{QH}
\end{equation}
where the contribution of UV ionization is
\begin{equation}
Q_{\rm H,UV}=\int_{\nu_{\rm H}}^{\nu_{\rm X}}\frac{L_{\nu}-L_{\nu}^{\rm abs}}{h_p\nu}d\nu,
\label{QHUV}
\end{equation}
while the contribution of X-ray ionization is
\begin{equation}
Q_{\rm H,X}\approx\int_{\nu_{\rm X}}^{\infty}\frac{L_{\nu}-L_{\nu}^{\rm abs}}{h_p\nu}d\nu+
\frac{1}{3}\int_{\nu_{\rm X}}^{\infty}\frac{L_{\nu}-L_{\nu}^{\rm abs}}{h_p\nu}
\left(\frac{\nu}{\nu_{\rm H}}-1\right)d\nu,
\label{QHX}
\end{equation}
where $\nu_{\rm H}$ is the frequency of photons with energy 13.6~eV; $\nu_{\rm X}$ 
is the frequency of the lowest energy X-ray photons (equal to 200~eV in this work);
The $1/3$ factor in the second term of the right hand side takes into
account the fact that, for a neutral gas, about one third of the energy of the high energy
electrons goes into the collisional 
ionization \citep{2004PhRvD..70d3502C, Valdes08}.
Eq. (\ref{QHX}) implicitly assumes that the collisional ionization is faster
than the production rate of free electrons by photons. 
This is always true if the density of the neutral material is 
sufficiently large, as in the case considered here.
If $N_{\rm H} = 1.5\times10^{25}~\rm cm^{-2}$, 
for BHs with mass $10^5$ ($10^6$)~$M_\odot$, $Q_{\rm H}\approx1.7\times10^{53}$ $s^{-1}$
($1.8\times10^{54}$ $s^{-1}$).

Under these conditions, the absorbed part of the spectrum, 
($L_\nu-L^{\rm abs}_\nu$), will finally escape 
from the Compton-thick material as the form of ``nebular emission", 
including the free-free, free-bound emission, and the two-photon emission.
The Ly$\alpha$ emission is not included here,
as the hydrogen density is large enough that the Ly$\alpha$ photons
are resonantly trapped in the surrounding gas
\citep{2011MNRAS.410.2025X,2006ApJ...652..902S,2010ApJ...712L..69S,2011MNRAS.411.1659L};
in this case the system is not able to lose energy through Ly$\alpha$ emission, instead,
essentially all the $n = 2$ to $n = 1$ transitions will eventually 
produce two-photon emission \citep{1970ApJ...160..939B,2010ApJ...712L..69S}.

The luminosity of free-free and the free-bound emission 
is simply $\propto Q_{\rm H}$. 
The expressions are \citep{2006ApJ...646..703F}
\begin{equation}
L^{\rm ff,fb}_{\nu}=4\pi\gamma_c^{\rm ff,fb}\frac{e^{-h_{\rm p}\nu/kT}}{\sqrt{T}}\frac{Q_{\rm H}}
{\alpha_{\rm B}(T)},
\label{Lff_fb}
\end{equation}
where $T$ is the temperature of the surrounding gas, $k$ is the Boltzmann constant, 
and $\alpha_{\rm B}$ is the case B recombination coefficient
\citep{1999ApJ...523L...1S}. The expression of $\gamma_c^{\rm ff,fb}$ is given by Eq. (12) of
\citet{2006ApJ...646..703F}. 

Atoms at $n = 2$ state is generated by both 
recombinations and collisional excitations, so the luminosity of two-photon emission is related to the 
rate of these two processes,
\begin{equation}
L_{\nu}^{\rm tp}=\frac{2h_{\rm p}\nu}{\nu_{{\rm Ly}\alpha}}P(\nu/\nu_{{\rm Ly}\alpha})
n_{\rm H}V[f^2_e
\alpha_{\rm B}(T)+f_e(1-f_e)C_{\rm coll}(T)],
\label{Ltp}
\end{equation}
where ${\nu_{{\rm Ly}\alpha}}$ is the frequency of Ly$\alpha$ photons, $P(\nu/\nu_{{\rm Ly}\alpha})$ 
is the normalized spectrum profile \citep{2006ApJ...646..703F}, $n_{\rm H}$ is the number density 
of the Compton-thick material while $V$ is the volume, $f_e$ is the ionization fraction and 
$C_{\rm coll}(T)$ is the collisional excitation rate \citep{2008ApJ...672...48C}.
Considering the two constraints: the energy conversion
$\int (L^{\rm ff}_\nu+L^{\rm fb}_\nu+L^{\rm tp}_\nu+L^{\rm abs}_\nu)d\nu = L_{\rm Edd}$
and the ionization equilibrium
$n_{\rm H}Vf_e^2\alpha_{\rm B}(T)=Q_{\rm H}$, to calculate the 
luminosity by Eqs. (\ref{Lff_fb} \& \ref{Ltp}),
we still need to know at least one of $n_{\rm H}V$, $T$ and $f_e$.
However, we find that with above two constraints,
the final nebular emission is very insensitive to the $f_e$, 
so in all our calculations we assume $f_e$ always equal to 0.5.

For $M_{\rm BH}=10^6~M_\odot$ and $N_{\rm H} = 1.5\times10^{25}~\rm cm^{-2}$,
we show the $L^{\rm ff}_\nu$, $L^{\rm fb}_\nu$, $L^{\rm tp}_\nu$ and $L^{\rm abs}_\nu$, 
and the final emerging spectrum (their sum) in the bottom panel of Fig. \ref{LQSO}. 
The $L^{\rm em}_\nu$ is the final spectrum we use for further calculations.
In the presence of Compton-thick material, 
the radiation between $\sim3 - 10$~eV, which will be redshifted into 
the near infrared bands at present day, is boosted by a factor of $\sim10$, 
while the emission in the X-ray bands 
is strongly suppressed. Similarly, UV photons $>$13.6 eV are completely absorbed so that these objects will 
not contribute to cosmic reionization.

\subsection{NIRB and CXB fluctuations }\label{formula}
The emissivity of a population of BHs formed at redshift $z$ with initial mass $M_{\rm BH, seed}$ accreting at the Eddington rate
for a time $t_{\rm QSO}$ and hosted in halos with 
virial temperature between $10^4$~K and $5\times10^4$~K is given by
\begin{equation}
\epsilon_{\nu}(z)=\frac{1}{4\pi}\int_{z_{\rm start}}^z
L^{\rm em}_\nu(M^\prime)\frac{dn_{\rm BH}}{dz^\prime}(z^\prime)dz^\prime,
\label{epsilon}
\end{equation}
where 
$M^\prime=M_{\rm BH,seed}{\rm exp}(t^\prime/t_{\rm Edd})$, $t^\prime$ is the time interval 
between $z$ and $z^\prime$,
$t_{\rm Edd} \approx 45$~Myr is the Eddington time scale
\citep{1964ApJ...140..796S,2007ApJ...665..107P}.
We assume that after the formation of a BH at redshift $z^\prime$,
the accretion only lasts for $t_{\rm QSO}$, so it can only radiate 
when $t^\prime < t_{\rm QSO}$.
We use a single typical value of $M_{\rm BH,seed}$ and $t_{\rm QSO}$
for all DCBHs. It is likely that the initial DCBH seeds span 
a mass range and are distributed according to a mass function $f(M_{\rm seed})$.
However,  the DCBH spectrum in the UV band (contributing to the present-day NIRB) 
is almost independent of the BH mass. Therefore, the DCBH contribution can be well represented by an average
mass $\langle M_{\rm seed}\rangle = \int M_{\rm seed}f(M_{\rm seed})dM_{\rm seed}/\int f(M_{\rm seed})dM_{\rm seed}$. 
In Eq. (\ref{epsilon}) the luminosity of a BH with mass $M^\prime$ 
$L_{\nu}^{\rm em}(M^\prime) = 0$ when $t^\prime > t_{\rm QSO}$.
The DCBH formation rate (per unit redshift) is
\begin{equation}
\frac{dn_{\rm BH}}{dz^\prime}=\int_{M_{\rm T4}}^{M_{\rm 5T4}}f_{\rm p}(M,z)\frac{d^2n}{dz^\prime dM}dM,
\end{equation} 
where $dn/dM$ is the halo mass function
\citep{1999MNRAS.308..119S,2001MNRAS.323....1S} and
$M_{\rm T4}$ and $M_{\rm 5T4}$ are the halo  masses corresponding to virial temperature
$10^4$~K and $5\times10^4$~K respectively. 
We assume the DCBH formation can not take place
in halos with virial temperature of $T_{\rm vir} \gtrsim 5\times 10^4$ K for
arguments given by \citet{2009MNRAS.396..343R}. 
The probability of a halo with mass $M$ to be still metal-free at redshift $z$, $f_{\rm p}$, 
is taken from \citet{2006MNRAS.369..825S}. 
As our redshift range is narrow, we neglect the evolution of 
$f_{\rm p}$ and use the results obtained for $z=15$.  
We use the $f_{\rm sn} = 0.1 $ model, see the 
red line of the upper right panel of Fig. 3 of \citet{2006MNRAS.369..825S}.
The value of $f_{\rm p}$ are $(0.9, 0.7, 0.4)$ for halos with virial
temperature of $(1, 2, 5)\times 10^{4}$~K.

Assuming the source BHs are in the redshift range [$z_{\rm start}$: $z_{\rm end}$],
with above emissivity, the cumulative flux of the NIRB we receive on Earth
is
\begin{equation}
\nu_0I_{\nu_0}=\nu_0\int_{z_{\rm start}}^{z_{\rm end}}\epsilon_{\nu}(z)\frac{dr_p}{dz}dz,
\label{nu0Inu0}
\end{equation}
where $r_p$ is the proper distance,
$\nu_0$ is the observed frequency, $\nu=(1+z)\nu_0$.

The contribution of accreting DCBHs to the angular power
spectrum of the NIRB can be computed by considering only the
two-halo correlation term, as each halo can host only one such object. The
BH power spectrum is then 

\begin{equation}
P_{\rm BH}(k,z)\approx P(k,z)b^2_{\rm eff}(z),
\end{equation}
where 
\begin{equation}
b_{\rm eff}(z)=\frac{1}{\bar{n}_h}\int_{M_{\rm T4}}^{M_{\rm 5T4}}b_h(M,z)\frac{dn}{dM}dM,
\label{beff}
\end{equation}
and $b_h$ is the bias of halos with mass $M$ relative to the matter
fluctuations \citep{2010ApJ...724..878T}  and $\bar{n}_h$ is the mean number density of halos,
\begin{equation}
\bar{n}_h = \int_{M_{\rm T4}}^{M_{\rm 5T4}}\frac{dn}{dM}dM.
\end{equation}
The angular power spectrum of the NIRB from BHs in the redshift
range [$z_{\rm start}$: $z_{\rm end}$] is \citep{2012ApJ...756...92C} 
\begin{equation}
C_l^{\rm NIRB} = \int_{z_{\rm end}}^{z_{\rm start}} dz \frac{[\nu\epsilon_{\nu}(z)e^{-\tau(\nu_0,z)}]^2}
{H(z)r^2(z)(1+z)^4}P_{\rm BH}(z),
\label{ClNIRB}
\end{equation}
where 
$r(z)$ is the comoving distance, and $H(z)$ is the Hubble parameter. The optical depth
$\tau(\nu_0,z)$, accounting for intergalactic absorption, is from \citet{2003MNRAS.339..973S}.

The same population, albeit Compton-thick, will also provide a
contribution to CXB fluctuations. This
can be computed with the above equations after replacing the term $\nu\epsilon_{\nu}(z)e^{-\tau} $
with $\int_{E_1(1+z)}^{E_2(1+z)}\epsilon_{E^\prime}(z)dE^\prime$ in
Eq. (\ref{ClNIRB}). Here, $E^\prime$ is the energy of photons with
frequency $\nu^\prime$,
$E_1$ and $E_2$ are the energy values delimiting the observed
X-ray band. We further assume that the IGM is transparent to X-rays
(i.e. $\tau(\nu_0,z)=0$ for X-rays).
Furthermore, DCBHs maybe sufficiently large to produce a detectable CXB-NIRB cross-correlation,
\begin{align}
&C_l^{\rm CXB-NIRB}= \nonumber \\
&\int_{z_{\rm end}}^{z_{\rm start}} dz \frac{[\nu\epsilon_{\nu}(z)e^{-\tau(\nu_0,z)}
\int_{E_1(1+z)}^{E_2(1+z)}\epsilon_{E^\prime}(z)dE^\prime]}
{H(z)r^2(z)(1+z)^4}P_{\rm BH}(z).
\end{align}

\begin{figure*}
\centering{
\subfigure{\includegraphics[scale=0.25]{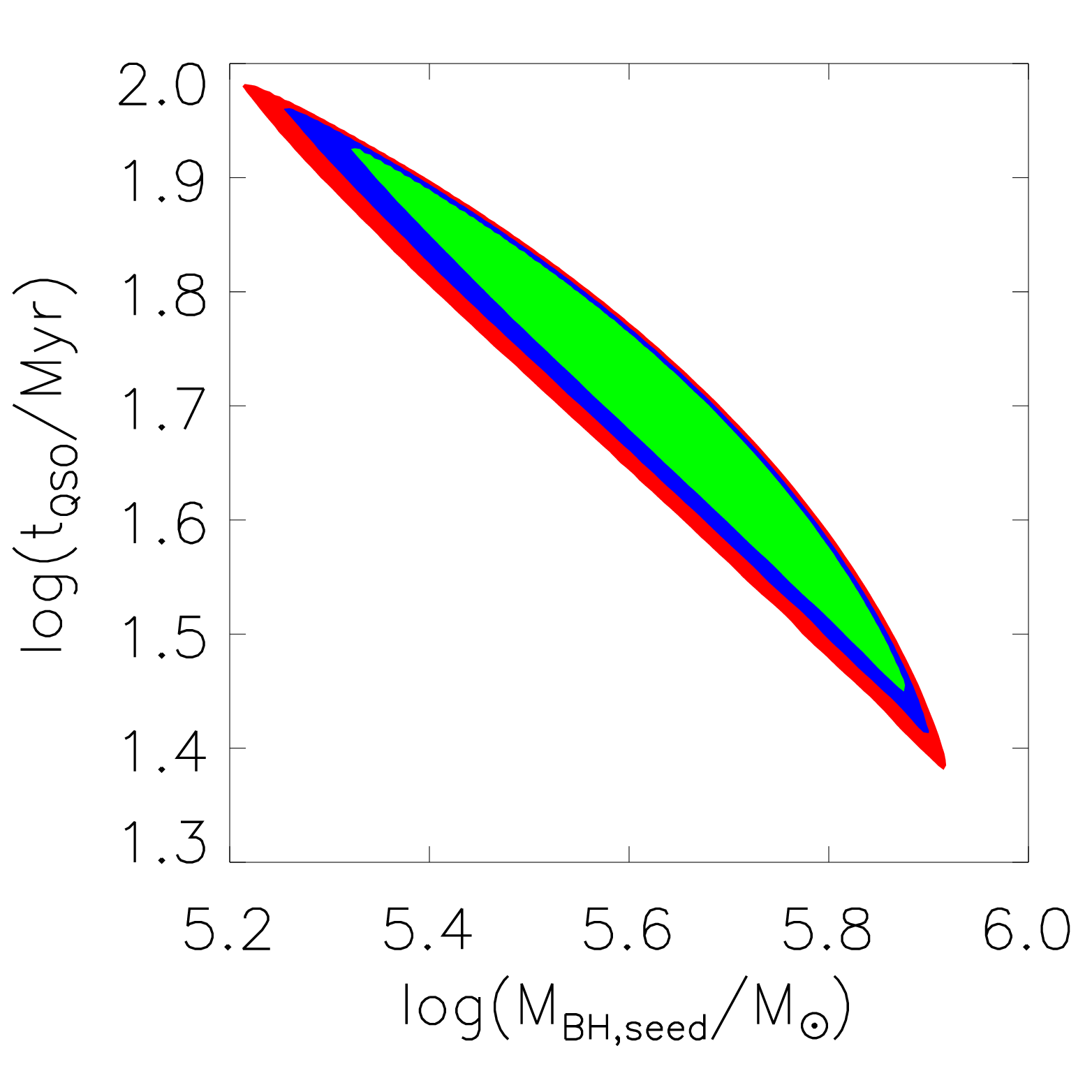}}
\subfigure{\includegraphics[scale=0.25]{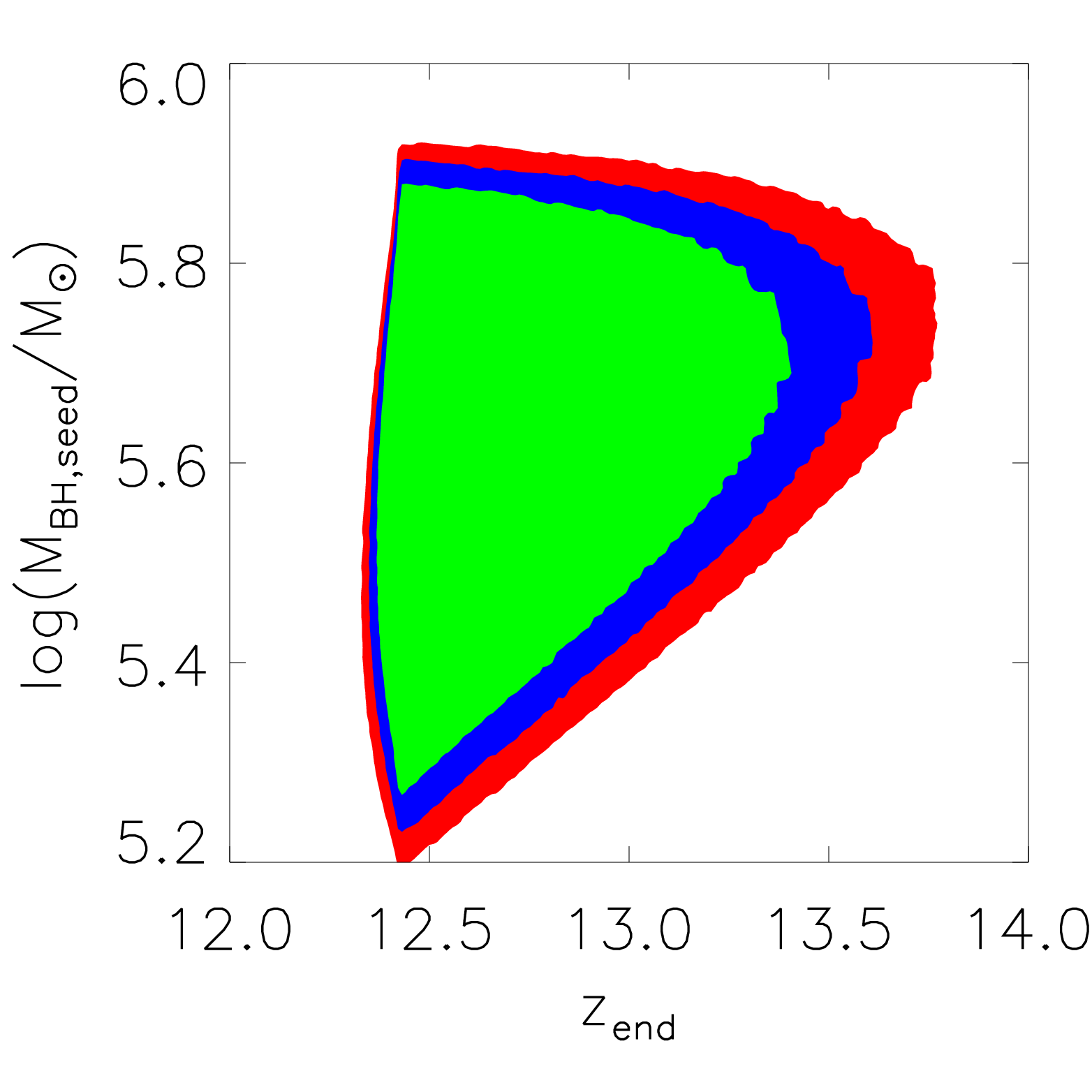}}
\subfigure{\includegraphics[scale=0.25]{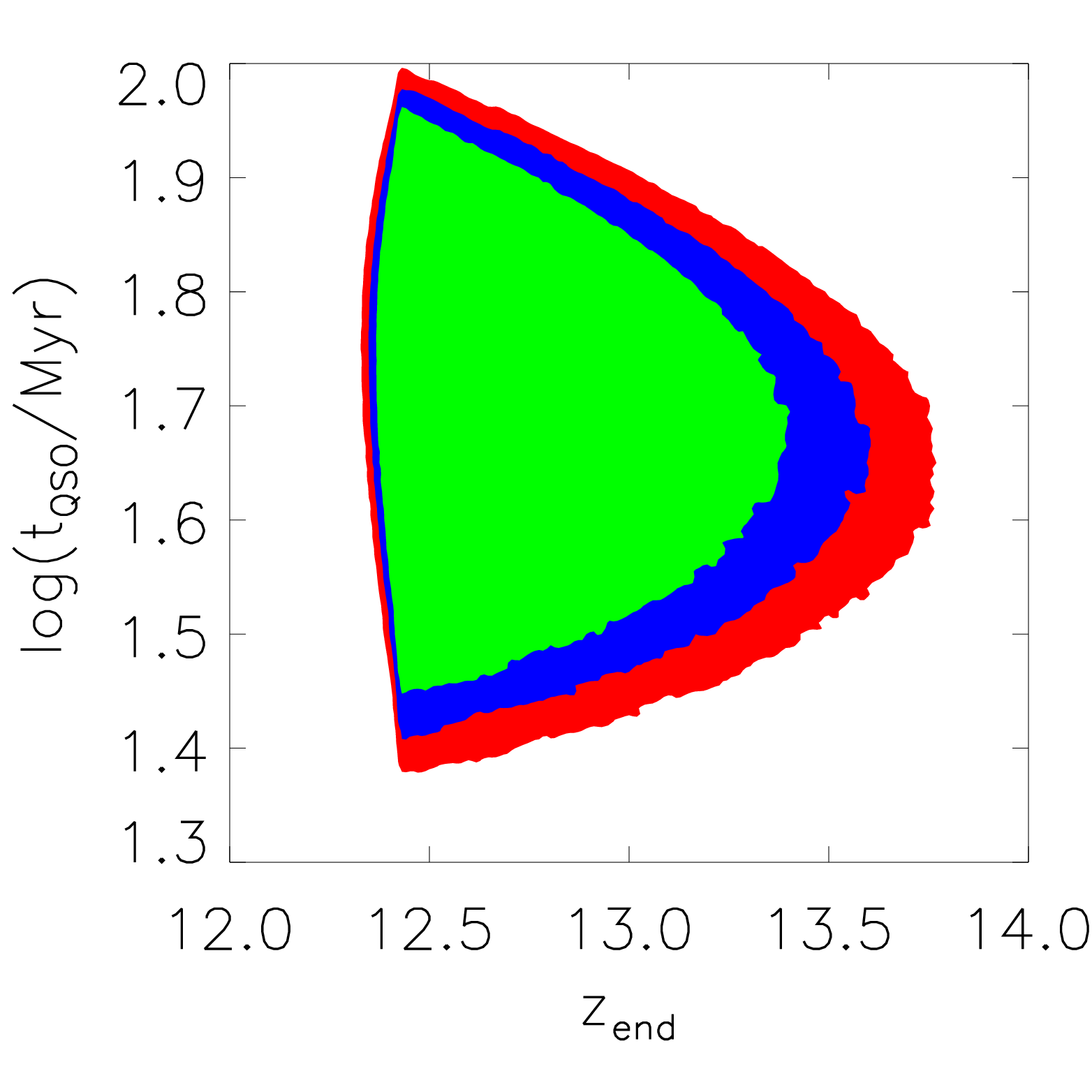}}
\caption{Constraints on the two dimensional parameters space of ${M_{\rm BH,seed}}-
t_{\rm QSO}$ (left),
$z_{\rm end}-M_{\rm BH,seed}$ (middle) and 
$z_{\rm end}-t_{\rm QSO}$ (right).
In each panel, regions filled by colors green, blue and red
correspond to 1 - 3$\sigma$ confidence level respectively.}
\label{contour}
}
\end{figure*}

\begin{figure}
\centering{
\includegraphics[scale=0.6]{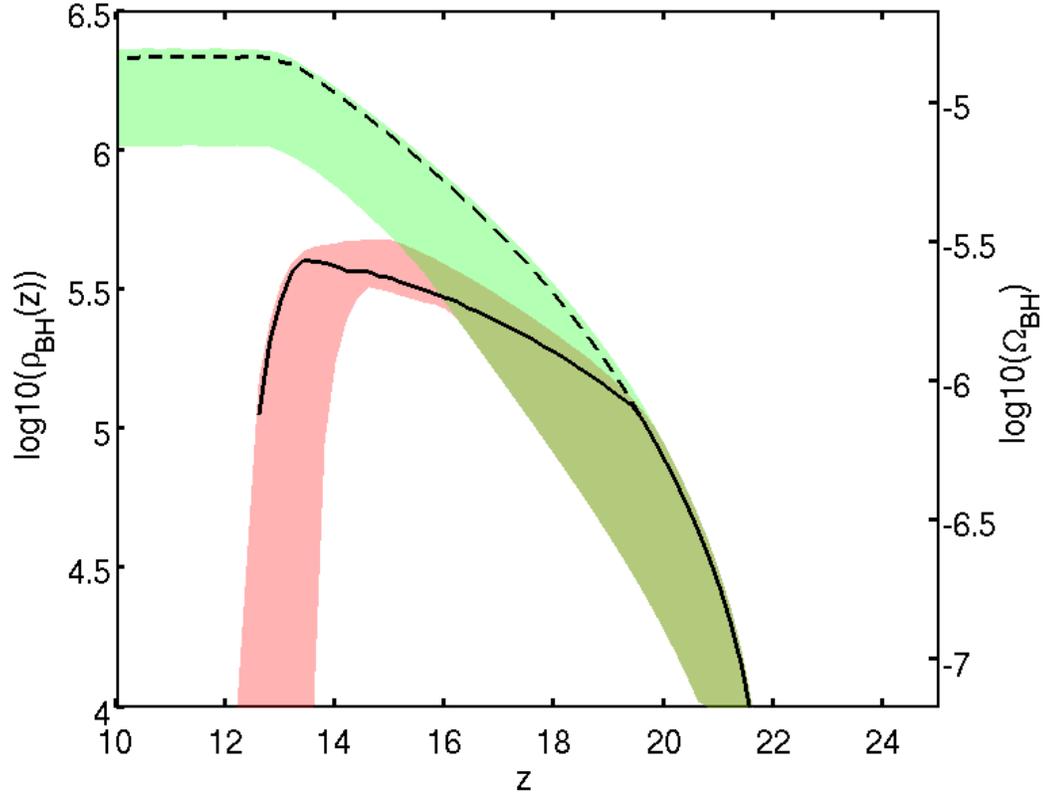}
\caption{
The cumulative mass density
(in units of $M_\odot$~Mpc$^{-3} $)
of DCBHs (dashed) and the accreting ones (solid),
only the latter contribute to the NIRB. 
Colored areas indicate regions corresponding to 1$\sigma$ dispersion around the
mean values of $(z_{\rm end}, M_{\rm BH,seed}, t_{\rm QSO})$.
The right $y$-axis gives $\Omega_{\rm BH}$, i.e., $\rho_{\rm BH}/\rho_c$, where $\rho_c$ the critical density.
}
\label{rhoBHz}
}
\end{figure}

\begin{figure}
\centering{
\includegraphics[scale=0.6]{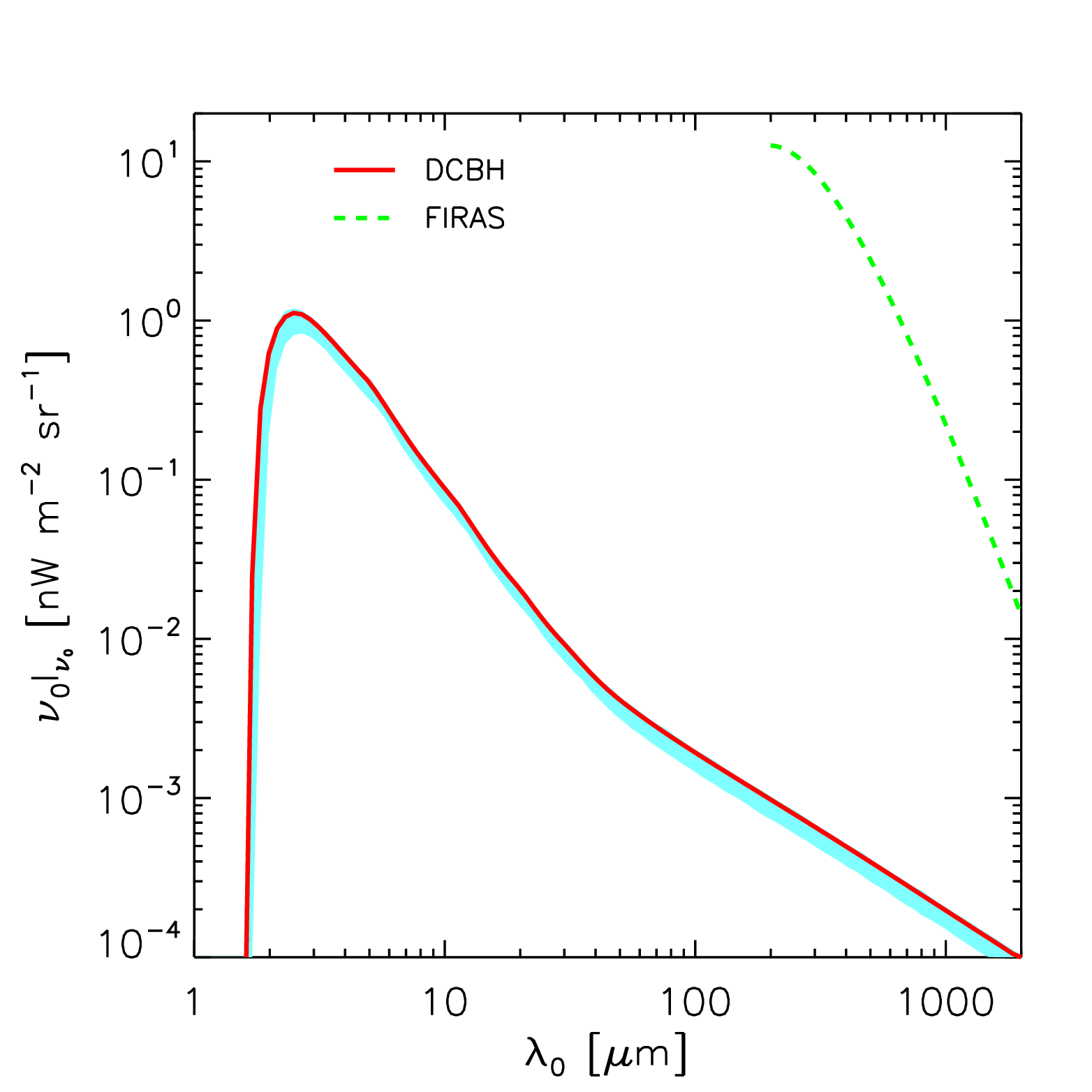}
\caption{
The contribution (solid) of DCBHs to the CIB with our best-fit parameters and 
the measured Cosmic Far-IR Background at 200-2000 $\rm \mu m$  (dashed) fitted by \citet{2000A&A...360....1G}.
Colored areas indicate regions corresponding to 1$\sigma$ dispersion around the
mean values of $(z_{\rm end}, M_{\rm BH,seed}, t_{\rm QSO})$.
}
\label{nu0Inu0Fig}
}
\end{figure}

DCBHs will also contribute to the unresolved fraction of the CXB flux. 
The analogous for the X-ray background flux seen 
by an observer at redshift $z=0$ and energy $E$ is given by
\begin{equation}
EJ(E)=E\int_{z_{\rm end}}^{z_{\rm start}}\epsilon_{E(1+z)}(z)\frac{dr_p}{dz}dz.
\end{equation}

\subsection{X-ray emission from undetected galaxies}
In addition to high redshift DCBHs,
it is necessary to include in the model also the contribution from low redshift galaxies
to the X-ray flux. In fact, high-mass X-ray binaries (HMXBs) dominate the X-ray emission of normal star-forming galaxies \citep{2012MNRAS.419.2095M,2011A&A...528A.149M}.
Their cumulative X-ray luminosity is found to correlate 
linearly with the total (obscured and unobscured) star-formation rate (SFR) of the galaxy 
\citep{2003A&A...399...39R,2012MNRAS.419.2095M}, so that
\begin{equation}
L_{\rm X,>0.5 keV}~({\rm erg~s^{-1}})=2.0\times10^{39}{\rm SFR}~(M_\odot~{\rm yr^{-1}}).
\label{LX}
\end{equation}
We can estimate the SFR in the UV, ${\rm SFR^0_{\rm UV}}$, from the galaxy UV luminosity by
\citep{2006ApJS..164...38I}
\begin{equation}
{\rm SFR}^0_{\rm UV}~(M_\odot~{\rm yr^{-1}})=1.2\times10^{-43}L_{\rm UV, obs}~({\rm erg~s^{-1}}),
\end{equation}
\noindent
where $L_{\rm UV, obs}$ is the observed cumulative UV luminosity 
at 2312\AA. In order to derive the total SFR when dust
obscuration is taken into account, we multiply ${\rm SFR^0_{\rm UV}}$ by a factor of 5.2\--- 
the typical value for $z\sim 2$ galaxies \citep{2012ApJ...744..154R} that
dominate the small scale NIRB angular fluctuations and the shot noise level
\citep{2012ApJ...752..113H}.
For an energy spectrum index 1.5, $L_{\rm X, 0.5-8 keV}$=0.75$L_{\rm X, > 0.5 keV}$,
so the factor we used in Eq. (\ref{LX}) is $\approx60\%$ of the measurement of 
\citet{2012MNRAS.419.2095M}.

With above formula, we can compute the contribution
of these faint galaxies to the CXB and the CXB-NIRB cross-correlation
by adopting the galaxy  number counts obtained in
\citet{2012ApJ...752..113H} and 
by simply replacing the near infrared flux of a galaxy with apparent magnitude $m$ with its X-ray flux.

\section{RESULTS}\label{results}

We compute the expected signal from a population of DCBHs formed at
high redshift. As already discussed, DCBHs form only during a limited redshift range.
Outside this range, the  DCBH formation is unlikely: at earlier times bright LW sources are very rare,
while later on the majority of atomic-cooling halos would have already been polluted by metals which 
induce a vigorous fragmentation
\citep{2002ApJ...571...30S}.  In the following we will discuss the
contribution of such DCBHs to the NIRB and to the CXB angular
fluctuations.

\subsection{NIRB fluctuations}

We fit the NIRB fluctuation data \citep{2012Natur.490..514C}  
by minimizing the $\chi^2$ with five free parameters, namely:  $z_{\rm
  start}$,~$z_{\rm end}$,  the mass of the DCBH seeds $M_{\rm
  BH,seed}$,  the accretion time scale $t_{\rm QSO}$, and the hydrogen
column density $N_{\rm H}$. 
However, we find that the 
best-fit $z_{\rm start}$ should be $\gtrsim 22$, but 
our results are insensitive
to the exact value of $z_{\rm start}$ as long as this is larger than
$z_{\rm start}=22$, as sources at higher redshifts contribute only little to the NIRB.
Therefore, in order to reduce the number of free parameters,
we fix to $z_{\rm start}=22$.
In addition,
as long as $N_{\rm H} \simgt 10^{24}~\rm cm^{-2}$,
the fit is also insensitive to this parameter,
because for such high neutral hydrogen column density,
the bulk of the emission $>$13.6 eV has already been re-processed 
to be nebular emission. We will show later on that a lower limit 
$N_{\rm H}>1.2-1.3\times10^{25}~\rm cm^{-2}$ is required not to exceed the 
CXB unresolved fraction. So in the following, we fix it to $N_{\rm
  H}=1.5\times10^{25}~\rm cm^{-2}$.

The fit has been performed by minimizing the $\chi^2$ over the three free parameters left and by requiring, as additional constraints, that the apparent magnitude of the single object should be $>30$ in the H-band of the HST/WFC3 and $>27$ in the 3.6 and 4.5~$\rm \mu m$ band of Spitzer/IRAC. Since at scales $\lesssim 100''$ the signal is dominated by the shot noise of low redshift, faint galaxies, the fit is performed to the large scale data only. We obtain the best-fit values and the one-parameter confidence interval: 
$z_{\rm end}=12.44^{+0.73}_{-0.07}$,
${\rm log}({M_{\rm BH,seed}}/{M_\odot})=5.85^{+0.03}_{-0.40}$ 
and ${\rm log}({t_{\rm QSO}}/{\rm Myr})=1.48^{+0.34}_{-0.03}$,
with a reduced $\chi^2_r=0.9$. A two-parameter confidence level study is presented in Fig.
\ref{contour}.
As shown in the left panel, there is degeneracy between the mass of the BH seeds and $t_{\rm QSO}$, 
as the contribution of a BH to the emissivity is approximately $\propto M_{\rm BH,seed}(e^{t_{\rm QSO}/t_{\rm Edd}}-1)$.
Even considering the redshift dependence, this degeneracy cannot be broken by using NIRB observations only. 
The sharp cut-off on  $z_{\rm end}$ at $\sim12.4$ present in the contour plots comes from
the assumption that individual DCBHs are not detected by current
instruments, and, in particular, by the fact that they should be fainter than $m = 30$
in the H-band of the HST/WFC3.
The sharp cut-off could be the result of our simplified assumptions on a
typical value of the seed mass and $t_{\rm QSO}$ and that a more realistic model
can result in a more gentle decline of $\rho_{\rm BH}$ at $z < 12.4$.

With above best-fit parameters, we calculate the density of all mass locked in BHs form
through direct collapse, and the density of DCBHs which are accreting 
at the corresponding redshift by
\begin{equation}
\rho_{\rm BH}(z)=\int_z^{z_{\rm start}}M^\prime\frac{dn_{\rm BH}}{dz^\prime}dz^\prime,
\end{equation}
when $t^\prime < t_{\rm QSO}$, 
$M^\prime=M_{\rm BH,seed}{\rm exp}(t^\prime/t_{\rm Edd})$;
when $t^\prime > t_{\rm QSO}$, $M^\prime=M_{\rm BH,seed}{\rm exp}(t_{\rm QSO}/t_{\rm Edd})$
for density of all DCBHs, while $M^\prime=0$ for accreting DCBHs.  
We plot these two densities
(and the BH density parameter $\Omega_{\rm BH}$ on right $y$-axis)
in Fig. \ref{rhoBHz}.
The shaded regions correspond to the $ 1\sigma$ uncertainty on
  the model free parameters.
Note that the upper limit of the uncertainty regions is also limited by 
our assumption that each BH is undetected in current surveys.
The surface density of the accreting DCBHs is 
$7.7\times10^6$~deg$^{-2}$ and their magnitudes at 
3.6 and 4.5 $\rm \mu m$ are below the current detection
limit of {\it Spitzer}.

The contribution of DCBHs to the background intensity, computed by
Eq. (\ref{nu0Inu0}), is shown in Fig. \ref{nu0Inu0Fig}; the
shaded region represent the 1$\sigma$  uncertainty on the model
free paramters.
In the NIR
bands this is about an order on magnitude below the one of ordinary
low-$z$ galaxies \citep{2012ApJ...752..113H} but higher than the
predicted signal of $z>6$ galaxies \citep{2013MNRAS.431..383Y}. At 3.6
and 4.5~$\mu$m bands DCBHs provide an intensity of 0.7 and 0.5 nW
m$^{-2}$ s$^{-1}$, respectively. 
The total contribution of undetected
sources (galaxies plus DCBHs) is still consistent with available
measures and limits on the unresolved NIRB fraction. At longer
wavelengths, the DCBH contribution declines rapidly to a value
$>100$ times smaller than the Cosmic Far-IR Background measured 
by FIRAS \citep{2000A&A...360....1G}.

\begin{figure*}
\begin{center}
\includegraphics[scale=0.4]{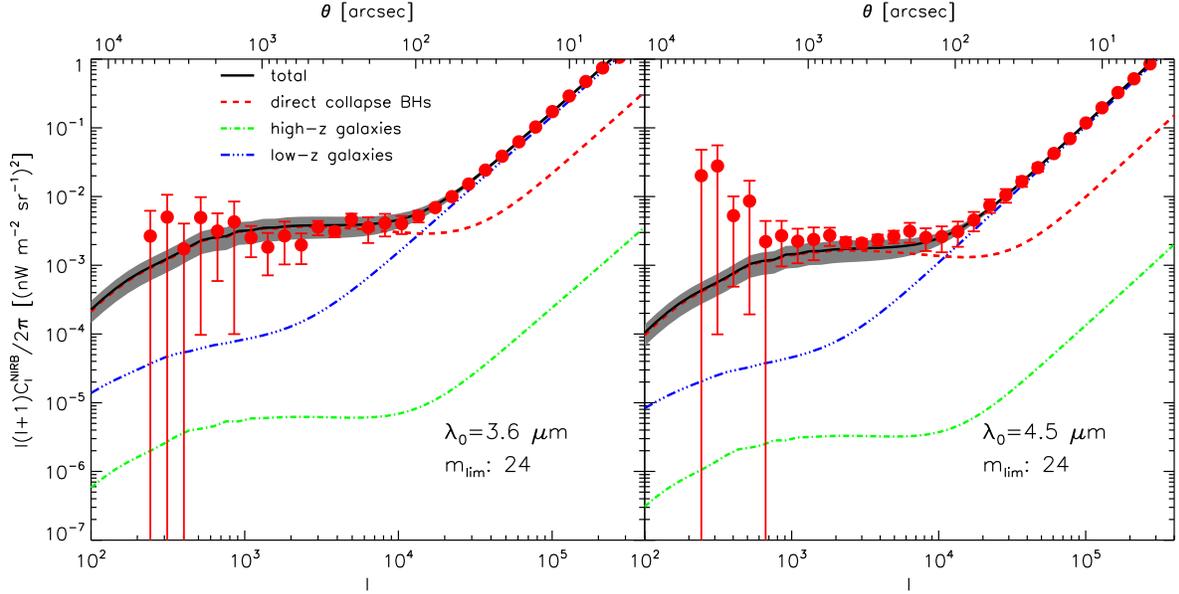}
\caption{NIRB angular power spectrum at wavelength 3.6~$\rm \mu m$ (left) 
and 4.5~$\rm \mu m$ (right), with the contribution from different sources: 
(a)  accreting DCBH (dashed line); (b) high redshift ($z > 5$) faint galaxies with 
$m>24$ (dot-dashed); (c) low redshift faint galaxies ($z<5$ and $m>24$,
dot-dot-dot-dashed) from \citet{2012ApJ...752..113H}. 
The solid line is the total with the shaded area marking the 1$\sigma$ goodness-of-fit.
Galaxy contributions are taken from \citet{2013MNRAS.431..383Y}.
The points are the latest measurements from \citet{2012Natur.490..514C}. 
The DCBH term accounts for the large scale (typically $> 100''$) 
clustering signal of the fluctuation spectra; at smaller scales shot
noise, i.e., Poisson fluctuations of source number counts in the beam, 
dominates.}
\label{NIRB}
\end{center}
\end{figure*}

We show the NIRB fluctuations produced by the population of high redshift DCBHs in Fig. \ref{NIRB} (dashed line).
We also plot the contribution from low redshift ($z < 5$) faint galaxies (dash-dotted-dotted-dotted)
by following the reconstruction of \citet{2012ApJ...752..113H}, and the contribution from high redshift 
($z > 5$) galaxies studied in \citet{2013MNRAS.431..383Y} (dash-dotted lines). The solid line is their 
sum, while the shaded regions represent the range of 1$\sigma$ goodness-of-fit.
Points with errorbars are measurements presented in \citet{2012Natur.490..514C}.
In the theoretical calculation we remove the galaxies brighter than $m =24$ to match the model 
shot noise level to the observed values.
As clearly seen in Fig. \ref{NIRB},  faint galaxies 
(including both the high redshift and low redshift ones) are unable to provide
the observed source-subtracted NIRB fluctuations at large scales,
which can be instead explained by the accreting DCBHs.

\begin{figure}
\begin{center}
\includegraphics[scale=0.6]{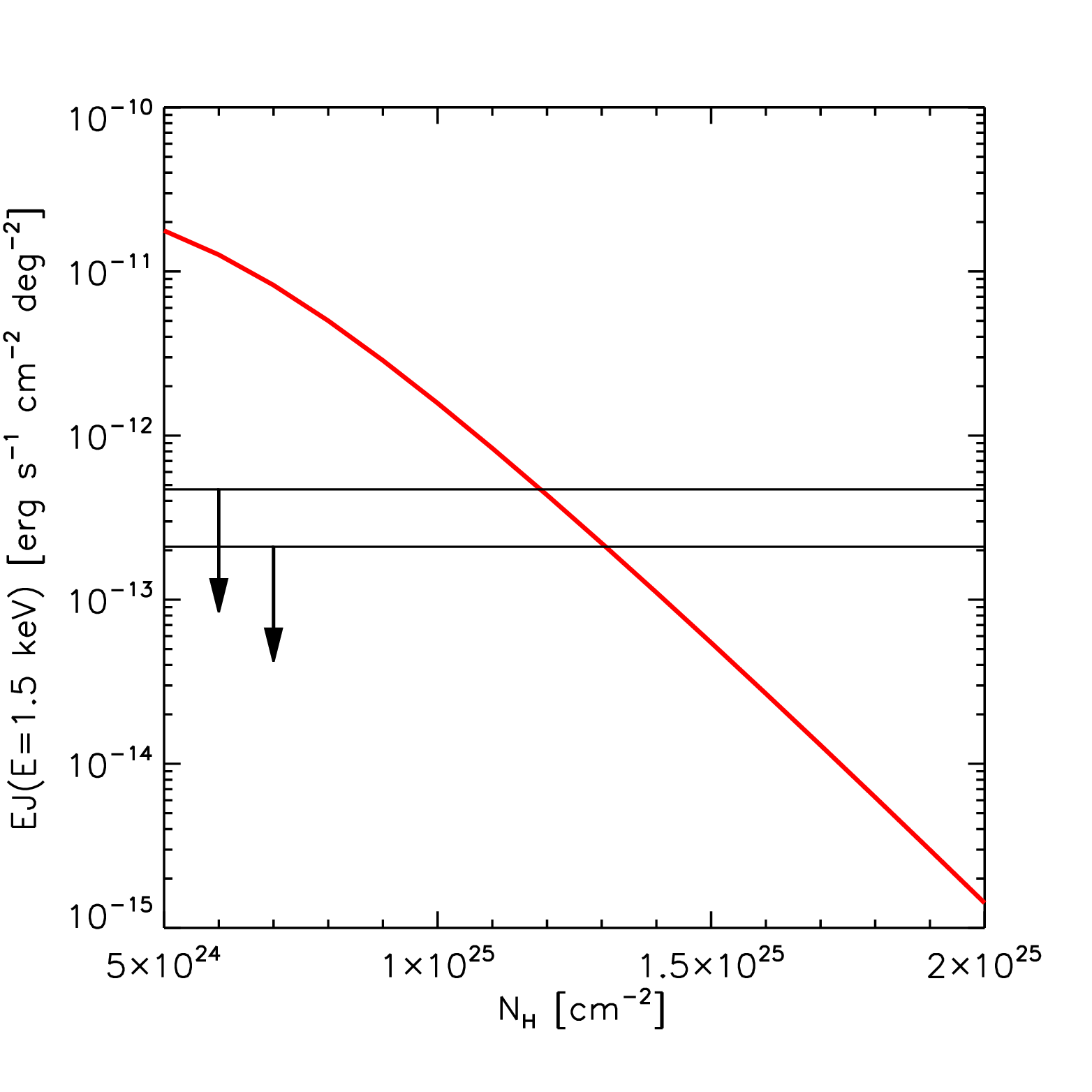}
\caption{Contribution of DCBHs to the CXB flux at 1.5 keV as a function of $N_{\rm H}$.
The two horizontal lines (marked by arrows) refer to the recent upper limits to the
unresolved fraction by \citet{2012A&A...548A..87M}: upper line is the maximum
value allowed by the data ($0.47\times10^{-12}~\rm
erg\,s^{-1}cm^{-2}deg^{-2}$), while the lower one
($0.21\times10^{-12}~\rm erg\,s^{-1}cm^{-2}deg^{-2}$)
is the more stringent limit obtained by subtracting the contributions of the
low redshift AGNs modeled by \citet{2007A&A...463...79G}.
Hydrogen column densities
$>1.2-1.3\times 10^{25}~\rm cm^{-2}$ are required to not exceed
the observational limits.}
\label{EJ}
\end{center}
\end{figure}

\begin{figure*}
\begin{center}
\includegraphics[scale=0.4]{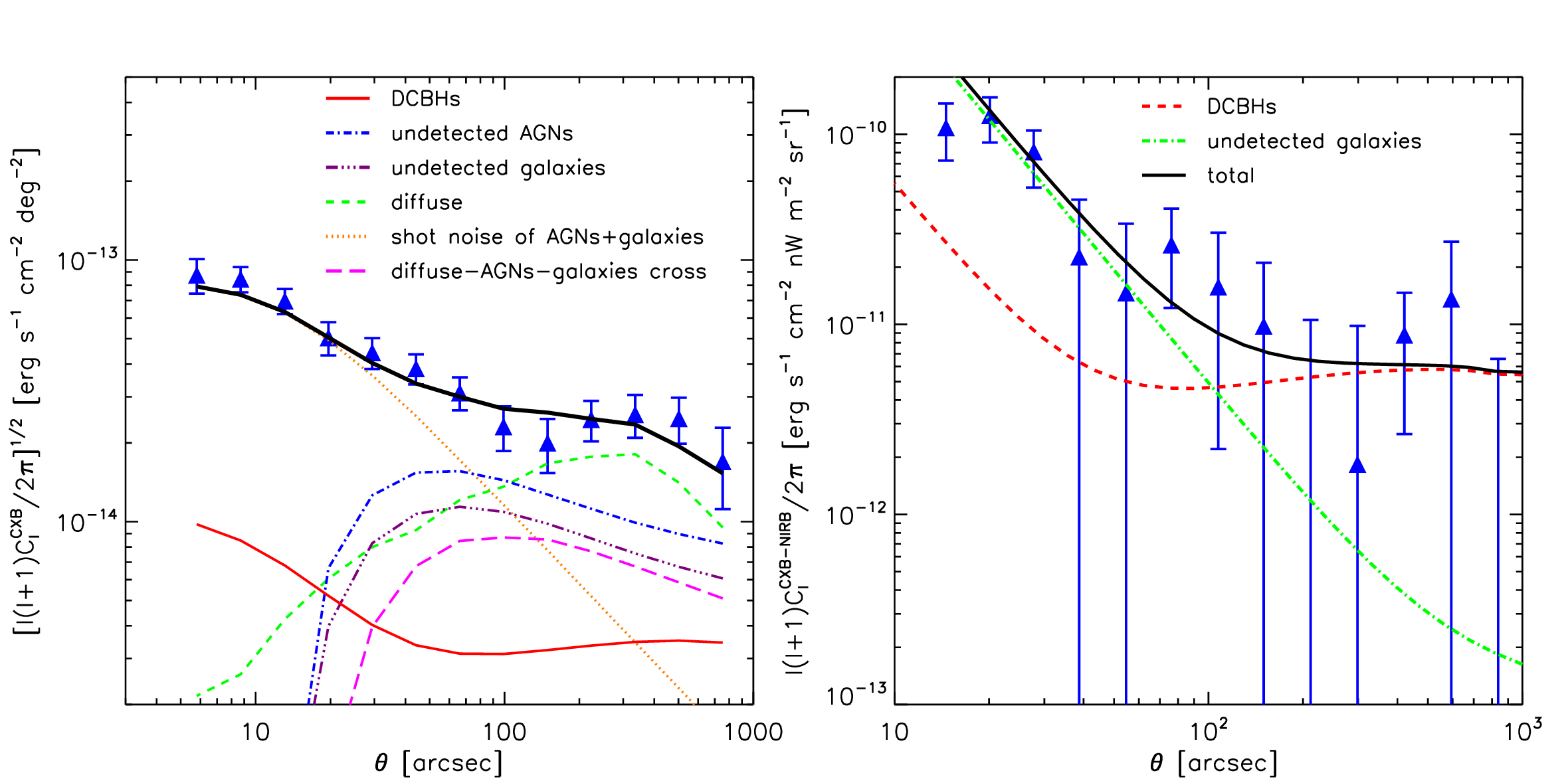}
\caption{\textit{Left panel:} Angular power spectrum of the CXB between 0.5-2.0~keV showing contribution from different
sources: DCBH (solid), undetected AGNs (dot-dashed), X-ray undetected galaxies (dot-dot-dot-dashed),
hot intergalactic medium (short-dashed), shot noise of AGNs and galaxies  (dotted). The solid thick line is the sum of 
different components. The data (triangles) and all curves except from the DCBHs are from
\citet{2012MNRAS.427..651C}. 
\textit{Right panel: } CXB (0.5-2.0~keV)-NIRB(4.5 $\mu$m)
cross-correlation power spectrum. The predicted signal from DCBHs is
shown with the dashed line for $N_{\rm H}=1.5\times10^{25}~\rm cm^{-2}$,
while the contribution from faint X-ray emitting galaxies is shown 
with dash-dotted line, the solid line is the sum. 
Data points are from \citet{2012arXiv1210.5302C}.
}
\label{CXB}
\end{center}
\end{figure*}

\subsection{CXB fluctuations and cross-correlation}

Albeit Compton-thick, DCBH may still contribute to the unresolved fraction of the CXB and to its 
spatial fluctuations. For the best-fit parameters of our model presented previously,
we try to vary the adopted value of the hydrogen column
density. We find that $N_{\rm H} >1.2-1.3\times10^{25}~\rm cm^{-2}$ is required 
(see Fig. \ref{EJ} where the CXB flux as a function of $N_{\rm H}$ is 
shown) not to exceed available limits \citep{2012A&A...548A..87M}, 
further supporting our assumption of
Compton-thick accretion. We note that this also 
reduces the tension between the need for an efficient and rapid growth
of BH seeds into the SMBHs powering quasar activity at $z=6-7$ and the strict
upper limit on the global accreted mass deduced
from X-ray background observations \citep{2012A&A...545L...6S}.

We also show the contribution to the angular power spectrum of the CXB 
from DCBHs in the left panel of Fig. \ref{CXB}, when 
$N_{\rm H}$ is set to be $1.5\times10^{25}~\rm cm^{-2}$,
together with the contributions from other components modeled 
in \citet{2012MNRAS.427..651C}, and the latest measurements in 
the same paper. The contribution of DCBHs to CXB fluctuations is found
to be negligible with respect the other sources at all angular scales.

As DCBHs contribute to both the NIRB and the CXB, it is natural to expect a CXB-NIRB
cross-correlation signal (dashed line in the right panel of Fig. \ref{CXB}). For $N_{\rm H}=1.5\times10^{25}~\rm
cm^{-2}$ the expected CXB (0.5-2 keV)-NIRB ($4.5 \rm \mu m$) cross-correlation signal is $\simeq 8\times
10^{-12}$ erg s$^{-1}$ cm$^{-2}$ nW m$^{-2}$ sr$^{-1}$ at scales $\theta> 100''$, as indeed tentatively 
reported by \cite{2012arXiv1210.5302C}.  At lower
scales, the CXB-NIRB is dominated by undetected low-$z$ galaxies
(dashed-dotted line). This has been computed as discussed in Section 2.3
adopting a limiting magnitude $\approx25$ 
to let the  model predicted shot noise level
of the NIRB match the observation
\citep{2012ApJ...753...63K,2012arXiv1210.5302C}. We also 
check that the corresponding X-ray flux is below the point source flux limit
of \citet{2012arXiv1210.5302C} observations\footnote{We note that the cross-correlation signal decreases for larger obscuring column densities; thus, in principle, the CXB-NIRB cross-correlation can be used to obtain a more accurate determination of 
$N_{\rm H}$ once more precise data will become available.}.

In conclusion, the observed CXB-NIRB cross-correlation can be explained by a
combination of X-ray emission from HMXBs in faint, low-$z$ galaxies at
small angular scales and the DCBH contribution at large scales.
We note that at very small scales, i.e., $\theta \simlt 20''$,
the observed cross-correlation drops. This is likely due to the instrument PSF of the X-ray 
observation, and (possibly) beam effects in NIRB observations, not modeled here.
The analogous cross-correlation for the CXB 2-7~keV band is found to be negligible, in agreement with
the findings of \citet{2012arXiv1210.5302C}.

\section{CONCLUSIONS}\label{conclusions}
We studied the spectrum of accreting black holes formed through 
the direct collapse of metal-free gas in halos with virial temperature $\simgt 10^4$~K. 
BHs formed by this process are very likely to be Compton-thick, so that:
(a) as most of photons with $h\nu >13.6$ eV are absorbed by the large column density of surrounding gas, 
the contribution of these objects to reionization is negligible and that to the CXB is 
reduced significantly; (b) ionizing photons are re-processed into optical-UV photons (free-free, free-bound 
and two-photon emission) while Ly$\alpha$ photons are trapped and finally converted into
two-photon emission. These secondary photons eventually escape the object and considerably boost 
(by a factor $\sim 10$) the contribution of these sources to the NIRB. 

We calculated the contribution of DCBHs to the NIRB fluctuations, by fitting the latest observations 
at 3.6 and 4.5~$\rm \mu m$. We find that observed fluctuations at angular scales larger than $\theta=100''$ 
can be explained by DCBHs formed in metal-free halos with virial temperature $T_{\rm vir}=1-5\times
10^4$~K down to $z_{\rm
  end}=12.44^{+0.73}_{-0.07}$, with initial masses 
${\rm log}({M_{\rm BH,seed}}/{M_\odot})=5.85^{+0.03}_{-0.40}$ and
accreting gas at the Eddigton limit for a time 
${\rm log}({t_{\rm QSO}}/{\rm Myr})=1.48^{+0.34}_{-0.03}$. 

Using the above best-fit parameters, we have calculated the DCBH contribution to the CXB
intensity at 1.5~keV, finding it  well below the current observational
limits as long as the sources are Compton-thick with $N_{\rm H}>1.2-1.3\times10^{25}~\rm cm^{-2}$. 
Analogously, DCBHs contribute only little to the CXB angular power spectrum. 

However, we predict that the DCBHs signal could emerge in the CXB-NIRB cross-correlation 
at scales $>100''$. For $N_{\rm  H}=1.5\times10^{25}~\rm cm^{-2}$ the cross-correlation level of the
DCBH population is $\simeq 8\times10^{-12}$ erg s$^{-1}$ cm$^{-2}$ nW m$^{-2}$ sr$^{-1}$,
in good agreement with recent observations, despite the remaining large uncertainties 
in current data. In addition, we also found that the observed cross-correlation signal 
at small scales ($<100''$) can be explained by the HMXBs hosted in faint, low-$z$
galaxies also dominating the small scale NIRB angular fluctuations.

Thus, the NIRB fluctuations and their cross-correlation with the CXB
might be the smoking gun of a peculiar population of early intermediate mass BHs; 
they might also shed light on the challenging questions posed by the rapid formation
of SMBHs seen in quasars.

\section*{ACKNOWLEDGMENTS}
It is a pleasure to acknowledge intense discussions and data exchange with N. Capelluti, A. Cooray, K. Helgason, E. Komatsu, T.
Matsumoto, R. Thompson, S. Kashlinsky, S. Mitra and T. Choudhury. We aknowledge financial support from PRIN MIUR 2010-2011, project ``The Chemical and Dynamical Evolution of the Milky Way and Local Group Galaxies'', prot. 2010LY5N2T. AF thanks UT Austin for support and hospitality as a Centennial B. Tinsley Professor and the stimulating atmosphere of the NIRB Workshop organized by the Texas Cosmology Center. BY thanks support and hospitality by SNS through the Distinguished Visiting Program. BY and XC also acknowledges the support of the NSFC grant 11073024 and the MoST Project 863 grant 2012AA121701. YX is supported by China Postdoctoral Science Foundation and by the Young Researcher Grant of National Astronomical Observatories, Chinese Academy of Sciences.

%\bibliography{refe}
%\bibliographystyle{mn2e}

\appendix

\section{Conditions for BH formation by direct collapse}\label{discussions}
The key factor for the formation of a BH by direct collapse is the destruction of $\rm H_2$ molecules
to prevent gas fragmentation. This is possible when the halo is immersed in a strong radiation field that  
either directly photo-dissociates $\rm H_2$ molecules via UV LW photons 
\citep{2001ApJ...548..509M,2008ApJ...673...14O}, or photo-detaches $\rm H^{-}$, the critical intermediate 
species for the $\rm H_2$ formation channel \citep{1997ApJ...474....1T,1997NewA....2..181A}, by photons  
with energies \gtsima0.76~eV \citep{2010MNRAS.402.1249S}. Relatively high values of the radiation field intensity
are required in order for $\rm H_2$ suppression to be effective, since for high gas densities self-shielding effects
\citep{2000ApJ...534...11H,1996ApJ...468..269D,2011MNRAS.418..838W} can protect $\rm H_2$. 
Therefore, at early times, the formation of DCBHs relies on  local fluctuations of the radiation field 
and it is possible only in biased regions where the background intensity from nearby sources is sufficiently large\footnote{This situation has been investigated by \citet{Dijkstra08} and \citet{Agarwal12} who discussed the case in which 
the LW is produced by a combination of normal galaxies and Pop III stars. They conclude that the DCBH formation rate 
is modest. For example, \citet{Agarwal12} gives
$\sim10^{-3}-10^{-2}~{\rm Mpc}^{-3}\,z^{-1}$. Their results do not conflict with our discussion because in their 
model radiation from DCBHs themselves was not included.}.  Once a massive BH eventually formed, it will also act as an additional source of UV photons, thus inducing a positive feedback. We assume the DCBH formation can not take place
in halos with virial temperature of $T_{\rm vir} \gtrsim 5\times 10^4$ K for
arguments given by \citet{2009MNRAS.396..343R}. 

\begin{figure}
\begin{center}
\includegraphics[scale=0.6]{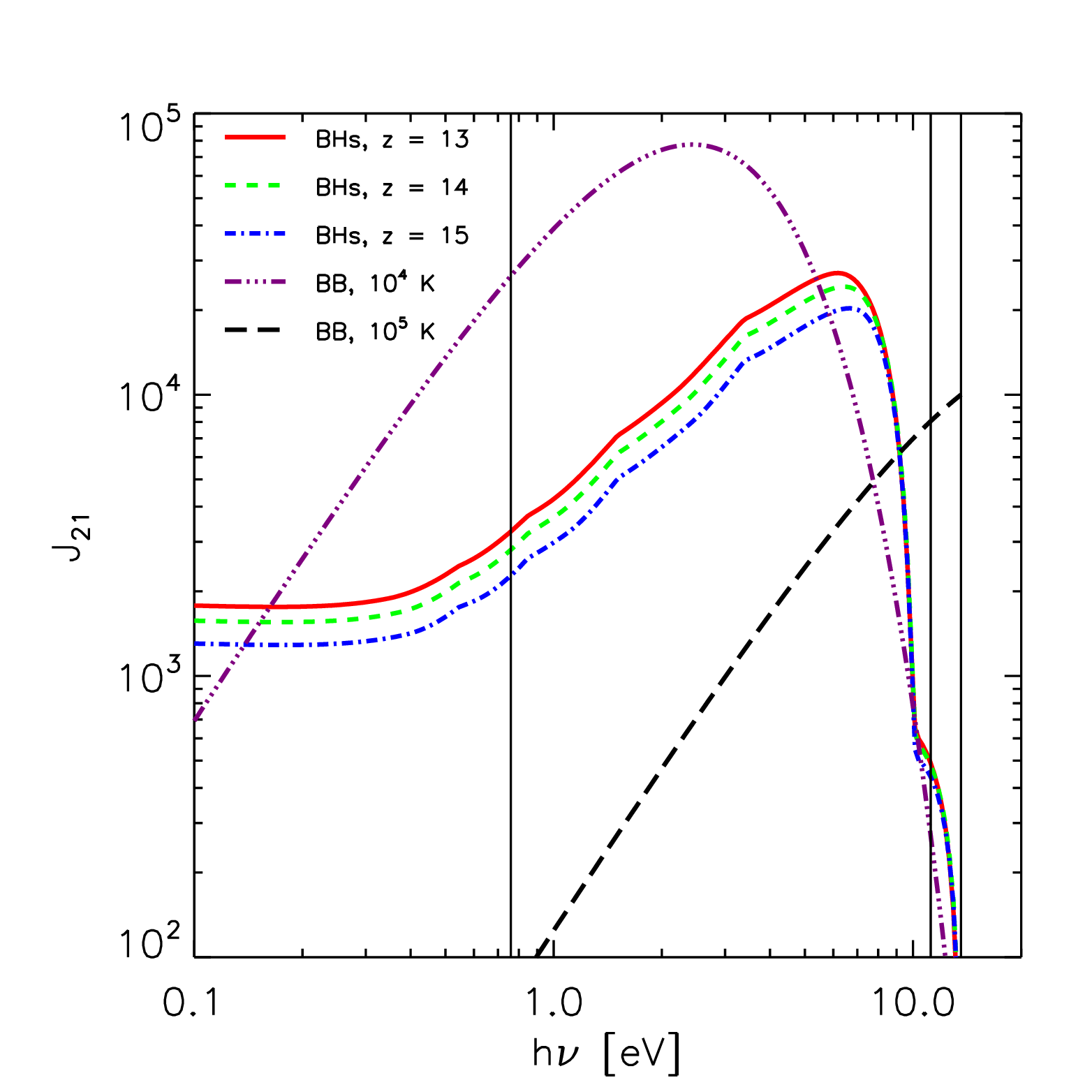}
\caption{ The background radiation field from the population of
accreting DCBHs between 0.1~eV and 13.6~eV computed at $z = 13, 14, 15$
and shown with the solid, dashed, dot-dashed lines, respectively. We also show the critical
spectrum for the formation of DCBHs from \citet{2010MNRAS.402.1249S} corresponding
to black body temperature $10^4$~K and $10^5$~K.
$J_{21}$ is the special intensity in units of $10^{-21}~\rm erg\,s^{-1} cm^{-2} Hz^{-1} sr^{-1}$.
For the former
$J_{21} = 30$ at
13.6~eV, while for the latter $J_{21}  = 10^4$
at 13.6~eV. By thin vertical lines,
we mark $h\nu = $0.76~eV, above which the
$\rm H^{-}$ dissociation is possible, and the interval $h\nu = 11.2~\rm eV - 13.6~\rm eV $, i.e., the
LW radiation range.}
\label{J21}
\end{center}
\end{figure}

At $z=15$ the intensity of the radiation of an accreting BH with mass 
$10^6$ M$_\odot$
at 12.87 eV drops below  $50\times10^{-21}~\rm erg\,s^{-1}cm^{-2}Hz^{-1}sr^{-1}$
at a distance $r\approx90$~kpc (comoving). Assuming that this radiation field is sufficient to suppress $\rm
H_2$ formation (the threshold value might depend on the detailed spectral shape, see below), the formation of another BH by 
direct collapse is possible in halos with $10^4~\rm K \lesssim T_{\rm vir} \lesssim 5\times10^4~\rm K$ hosted in a 
region between about two virial radii of the first halo and $r$. The number of neighbors in this region can be computed from  
the two-point correlation function $\xi(r,z)$,
\begin{equation}
N_{\rm neig}(z) = \bar{n}_h\int_{r_{\rm min}}^{r_{\rm max}}
[1+\xi(r,z)]4\pi r^2dr.
\end{equation}
The two-point correlation function for a region of mass $M_1$ and linear overdensity $\delta_1$ is 
\begin{equation}
\xi(r,z)=\frac{1}{2\pi^2}\int P_{1,\rm BH}(k,z)\frac{\sin(kr)}{kr}k^2dk,
\end{equation}
where $P_{1,\rm BH}(k,z)$ is the power spectrum of the spatial fluctuations of BHs in this region. 
This is computed as $P_{1,\rm BH}(k,z)=P(k,z)b^2_{1,\rm eff}(z)$, where $P(k,z)$ is the mean cosmic matter 
power spectrum; the bias for this region, $b_{1,\rm eff}(z)$, is derived from Eq. (\ref{beff}) by 
replacing $dn/dM$ with the mass function of halos in the given region $M_1$, as computed by the 
Extended Press-Schechter theory \citep{1993MNRAS.262..627L,1996MNRAS.282..347M,2002MNRAS.329...61S}.

The above argument can be expressed in terms of a critical linear overdensity $\delta_c$ for which one 
neighbor is within the influence radius $r$ of the halo hosting the accreting BH. Regions with
$\delta_1>\delta_c$ have one or more neighbors and the BH can promote further BH formation by direct 
collapse in these halos.
The distribution of regions above the critical overdensity 
as a function of $M_1$ can be determined by the distribution of distances that 
random walks have traveled before first up-crossing the density barrier, 
and can be computed by using the method 
proposed by \citet{2006ApJ...641..641Z}. We find that at $z = 15$ in about 20\% of the cosmic 
volume, halos with $10^4~\rm K \lesssim T_{\rm vir} \lesssim 5\times10^4~\rm K$  have on average 
at least one neighbor in the same mass range within a radius of $\le 90$~kpc. 

The formation of the first DCBH must be triggered by the UV background produced by normal or 
Population III star forming galaxies. However, the collective radiation from newly formed BHs adds to this background
and further increases it. In Fig. \ref{J21} 
we show the BH specific background intensity $J(\nu)$ at frequency $\nu$
using the emissivity $\epsilon_{\nu}(z)$ predicted by our model and computed using the 
following expression:
\begin{equation}
J(\nu,z)=(1+z)^3\int_z^{z_{\rm start}}\epsilon_{\nu^\prime}(z^\prime){\rm e}^{-\tau}\frac{dr_p}{dz^\prime}dz^\prime,
\end{equation}
where $\nu^\prime=\nu(1+z^\prime)/(1+z)$, and $r_p$ is the proper distance. The precise $\rm H_2$ suppression intensity threshold for this specific spectral shape is unavailable as previous studies have only explored black-body (stellar) spectra of temperatures of $10^4$~K and $10^5$~K
\citep{2010MNRAS.402.1249S}, or a power law form spectrum
\citep{2001ApJ...546..635O} . The critical curves for these two black-body cases are also reported in 
Fig. \ref{J21} as a function of energy.

In the LW regime, our background radiation is weaker than the $10^5$~K black body critical spectrum,
but higher than the $10^4$~K one. On the other hand, for energies $\simgt 0.76$~eV, radiation from 
accreting DCBHs is much larger than the $10^5$~K critical case, yet lower than the $10^4$~K one. 
Thus, our spectrum is somewhat ``intermediate'' between those studied by \cite{2010MNRAS.402.1249S}. 
Lacking specific numerical simulations, it is presently hard to quantify the exact intensity threshold 
value; the previous results can only provide a consistency check.

\section{Photoelectric absorption and Compton scattering }\label{Y97}

In this Appendix, we briefly summarize the formulae derived by Y97 and used here to calculate the absorbed 
and scattered spectrum of a BH. This emerging spectrum is the sum of the unabsorbed primary spectrum 
and the flux from all scattering orders:

\begin{equation}
N(E)=N_{\rm i}(E)e^{-\tau_0}+\sum_1^{n_{\rm max}}N_n(E),
\end{equation}
where $N(E)$ is the emerging photon number flux at energy $E$, $N_{\rm i}(E)$ is the primary spectrum (corresponding to 
the solid line in the upper panel of Fig. \ref{LQSO}), while $N_n(E)$ is the flux of photons that have already been
scattered exactly $n$ ($n \le n_{\rm max}$) times and then escape the medium. We use $n_{\rm max} = 20$ after checking convergence 
of the results for $N_{\rm H}$ considered in this work. The optical depth $\tau_0$ is the sum of scattering optical depth and the 
photoelectric absorption optical depth,
\begin{equation}
\tau_0=\tau_{\rm s}+\sigma_{\rm abs}(E)N_{\rm H}=[1.2\sigma_{\rm T}+\sigma_{\rm abs}(E)]N_{\rm H},
\end{equation}
where $\sigma_{\rm T} = 0.67\times10^{-24}~\rm cm^{-2}$ is the cross-section 
of Thomson scattering. $\sigma_{\rm abs} (E)$	is the cross-section of 
photon ionization, when 0.03 keV $< E <$ 10 keV, the data is fitted by 
\citet{1992ApJ...400..699B}; for $E > 10$~keV, we get this 
cross-section by the extrapolation of a $\propto E^{-3}$ law. 

To get the final expression for $N_n(E)$ we proceed as follows. 
The fraction of photons  with initial energy $E_0$ that have never been absorbed or 
scattered by the medium is $P_0=e^{-\tau_0}$, so that a fraction $1-P_0$ of them
is either absorbed or scattered. Suppose a fraction $\lambda_0$ of these photons 
is scattered (rather than absorbed) by the medium, and after this scattering $G_1$ of them 
escape from the system; one then gets for the fraction of photons that escape the medium 
after one scattering, $P_1=(1-P_0)\lambda_0G_1$. Among the remaining part, 
$\lambda_1$ of them experience a second scattering and $G_2$ of them escape the medium 
after this scattering, so $P_2=[(1-P_0)\lambda_0(1-G_1)]\lambda_1G_2$ of the photons escape the 
medium after two scatterings, and so on. We finally get the complete expression for 
the series of $P_n$ (Eq. (6) of Y97)
\begin{equation}
P_n=\left(\prod_{i=0}^{n-1}\lambda_i-\sum_{j=0}^{n-1}P_j\prod_{i=j}^{n-1}\lambda_i\right)G_n.
\label{Pn}
\end{equation}

For photons with initial energy $E_0$, define a dimensionless wavelength $y_0=m_ec^2/E_0$, where 
$m_e$ is the mass of the electron and $c$ is the speed of light.
After $n$ scatterings, if the fraction of the scattered photons have wavelength 
between $y$ and $y+dy$ is 
$F_n(y,y_0)dy$, then the energy distribution of photons that escape the 
medium after $n$ scatterings is
\begin{equation}
N_n(E)=\int_{E}^{E_{\rm max}} dE_0N_{\rm i}(E_0)P_nF_n(y,y_0)\frac{dy}{dE}.
\end{equation}
$E_{\rm max}$ is an upper limit set by the fact that photons with energy above this limit cannot 
reach energy $E$ after $n$ scatterings. It is  convenient to change the integration 
variable $E_0$ into $y_0$; in this case we get
(Eq. (7) in Y97),
\begin{equation}
N_n(E)= \frac{m_e^2c^4}{E^2}\int_{y-2n}^{y}P_n
F_n(y,y_0)N_{\rm i}\left(\frac{m_ec^2}{y_0}\right){y_0}^{-2}dy_0,
\label{Nn}
\end{equation}
where $y-2n$ corresponds to the maximum energy that the initial photons contribute to 
photons with energy $E$, and it is determined by the fact that the maximum wavelength change
per scattering is 2 (Y97).

The remaining task is then to determine the
expressions of $F_n$ and $P_n$ (or $\lambda_n$ and $G_n$) in 
Eq. (\ref{Nn}).
The non-relativistic approximation of $F_n$ could be found in 
Sec. A.1 of \citet{1995MNRAS.272..481B}: 
\begin{equation}
F_1=\frac{3}{8}[1+(\Delta y -1)^2],
\end{equation}
where $\Delta y = y - y_0$;
\begin{equation}
F_2= \left\{
\begin{tabular}{ll}
$f(\Delta y)~~~~~~~~~~~~~~~~~ 0 \le \Delta y < 2$,\\
$f(4-\Delta y)~~~~~~~~~~~~~ 2 \le \Delta y < 4$, \\
0 ~~~~~~~~~~~~~~~~~~~~~~{\rm otherwise},
\end{tabular}
\right.
\end{equation}
where $f(\Delta y)$ is expressed as 
\begin{equation}
f(\Delta y)=\left(\frac{3}{8}\right)^2[4\Delta y-4(\Delta y)^2+2(\Delta y)^3-(\Delta y)^4/3+(\Delta y)^5/30];
\end{equation}
and 
\begin{equation}
F_n=\left(\frac{4\pi n}{5}\right)^{-1/2}{\rm exp}\left[\frac{-5(\Delta y-n)^2}{4n}\right]~~~n \ge 3.
\end{equation}			

The single-scattering albedo of photons with energy $E$ is 
$\lambda_0(E)=\tau_{\rm s}/\tau_0$. For photons with initial 
energy $E_0$, their wavelength distribution is $F_n(y,y_0)$ 
after they are scattered $n$ times by the medium. The albedo $\lambda_n$ could then 
be represented by the mean of $\lambda_0$ weighted by this 
wavelength distribution (Eq. (5) of Y97)
\begin{equation}
\lambda_n(E_0)=\int_{y_0}^{y_0+2n}\lambda_0(m_ec^2/y)F_n(y,y_0)dy.
\end{equation}

The series of $P_n$ in Eq. (\ref{Nn}) are obtained by recursion. 
First, the approximated expression of $P_1$ that fits the Monte Carlo results is  (Eq. (10)
of Y97)
\begin{equation}
P_1=\frac{A}{16J_1(\lambda_1,E_0)B},
\end{equation}
where $A=\tau_{\rm s}[4e^{-\tau_1}+3e^{-\tau_0}+2e^{-1/2(\tau_0+3\tau_1)}+2e^{-1/2(\tau_0+\tau_1)}
+4e^{-1/2(\tau_0+\sqrt{3}\tau_1)}+e^{-(\tau_0+2\tau_1)}]$,
$B=1+9.8600\times10^{-2}\tau_0-2.8717\times10^{-4}\tau_0^2+7.0954\times10^{-7}\tau_0^3$,	
and $\tau_1=\tau_{\rm s}/\lambda_1$.
$J_1$ is the $n =1$ case of the series of 
(Eq. (11) of Y97)
\begin{equation}
J_n(\lambda_n,E_0)=e^{-\tau_{\rm s}}+(1-e^{-\tau_{\rm s}})\lambda_n^{1/2(n-n_0+1)}
\end{equation}
when $n\ge n_0~\&~E_0 > E_{\rm K}$,
where $n_0={\rm max}\{{\rm int}[(E_{\rm K}^{-1}-E^{-1}_0)m_ec^2],0\}$,
and $E_{\rm K}$ is the Fe K edge energies; $J_n = 1$ otherwise.
$G_1$ is therefore determined according to Eq. (\ref{Pn}), 
\begin{equation}
G_1=\frac{P_1}{\lambda_0(1-e^{-\tau_0})}, 
\end{equation}
and the expressions of the series of $G_n$ are (Eq. (12) of Y97)
\begin{equation}
G_n=\frac{G_1[1+e^{f(\tau^\prime_n)}]}{J_n(\lambda_n,E_0)},
\end{equation}
where $\tau^\prime_n=\tau_{\rm s}\lambda_{n-1}/(\lambda_0\lambda_{n-2})$,
$f(\tau^\prime_n) = \alpha_n+\beta_ne^{-\tau^\prime_n/\gamma_n}+
(0.25+0.75e^{-\tau^\prime_n/\gamma_n}){\rm ln}\tau^\prime_n$ is fitted 
to the Monte Carlo results.
For $n=2$ to 9, parameters $\alpha_n$, $\beta_n$ and $\gamma_n$ are listed in Table 1
of Y97, and $G_{>9} = G_9$. 

\end{document}